# On Irreversibility, Dissipation and Response Theory

Denis J. Evans,[1] and Debra J. Searles[2]

[1] Research School of Chemistry, Australian National University, Canberra, ACT 0200, Australia

[2] Nanoscale Science and Technology Centre, School of Science, Griffith University, Brisbane, Qld 4111 Australia

**Abstract**

Recently there has been considerable interest in the Fluctuation Theorem (FT). The FT shows how time reversible microscopic dynamics leads to irreversible macroscopic behavior as the system size or observation time increases. We show that the argument of the Evans-Searles FT, the dissipation function, plays a central role in nonlinear response theory and derive the Dissipation Theorem, giving exact relations for nonlinear response of classical N-body systems. These expressions should be verifiable experimentally. When linearized they reduce to the Green-Kubo expressions for linear response.

PACS: 05.70.Ln, 05.40.-a, 05.20.-y, 73.50.-h

Consider a classical system of N interacting particles in a volume V. The microscopic state of the system is represented by a phase space vector of the coordinates and momenta of all the particles, in phase space - $\{\mathbf{q}_1,\mathbf{q}_2,..\mathbf{q}_N,\mathbf{p}_1,..\mathbf{p}_N\} \equiv (\mathbf{q},\mathbf{p}) \equiv \Gamma$ where $\mathbf{q}_i, \mathbf{p}_i$ are the position and momentum of particle i. Initially (at t = 0), the microstates of the system are distributed according to a normalized probability distribution function $f(\Gamma,0)$. We separate the N particle system into a system of interest and a reservoir region containing $N_W$ particles. We shall assume that the reservoir region contains many more particles than the system of interest, $N_W >> (N - N_W)$, and we write the equations of motion for the composite N-particle system, as,

$$\dot{\mathbf{q}}_i = \mathbf{P}_i/m + \mathbf{C}_i(\Gamma) \cdot \mathbf{F}_e$$
$$\dot{\mathbf{p}}_i = \mathbf{F}_i(\mathbf{q}) + \mathbf{D}_i(\Gamma) \cdot \mathbf{F}_e - S_i \alpha(\Gamma) \mathbf{p}_i$$
(1)

where $\mathbf{F}_e$ is the dissipative external field that couples to the system via the phase functions $\mathbf{C}_i(\Gamma)$ and $\mathbf{D}_i(\Gamma)$, $\mathbf{F}_i(\mathbf{q}) = -\partial \Phi(\mathbf{q})/\partial \mathbf{q}_i$ is the interatomic force on particle i, and $\Phi(\mathbf{q})$ is the interparticle potential energy, and the last term $-S_i \alpha(\Gamma) \mathbf{p}_i$ is a deterministic time reversible thermostat used to add or remove heat from the particles in the reservoir region. [1-3] The thermostat multiplier is chosen using Gauss' Principle of Least Constraint [1-3], to fix some thermodynamic constraint (e.g. temperature or energy). The thermostat employs a switch, $S_i$, which controls how many and which particles are thermostatted, $S_i = 0$;  $1 < i < (N - N_{therm})$ , $S_i = 1; (N - N_{therm} + 1) \leq i \leq N, N_{therm} \leq N_W$. The equations of motion for the particles in the system of interest are quite natural. The reservoir region is assumed to not interact with the dissipative field $\mathbf{C}_i, \mathbf{D}_i = \mathbf{0}; (N - N_W + 1) \leq i \leq N$ and the equations of motion for the more distant reservoir particles, $N_W < N_{therm} \leq i \leq N$, are supplemented with the unnatural thermostat term. It is worth pointing out that as described, equations (1) are time reversible and heat can be both absorbed and given out by the thermostat.

One should not confuse a real thermostat composed of a very large (in principle, infinite) number of particles with the purely mathematical -albeit convenient- term $\alpha$. In writing (1) it is assumed that the reservoir momenta $\mathbf{p}_i$ are peculiar (i.e. measured relative to the local streaming velocity of the fluid or wall). The thermostat multiplier is chosen to fix the peculiar kinetic energy of the wall particles

$$K_{therm} \equiv \sum_{S_i=1} p_i^2 / 2m = d_C N_{therm} k_B T_W / 2, \qquad (2)$$

with $N_{therm} = \sum S_i$. The quantity $T_W$ defined by this relation is called the kinetic temperature of the wall, and $d_C$ is the Cartesian dimension of the system. It is assumed that $N_W, N_{therm} \gg (N - N_{therm}) > (N - N_W)$. This means that the entire wall region can be assumed to be arbitrarily close to equilibrium at the thermodynamic temperature $T_W$.

One might object that our analysis is compromised by our use of these artificial (time reversible) thermostats. However the artificial thermostat region can be made arbitrarily remote from the system of interest by ensuring that the particles with $S_i = 1$ are far from the system of interest [4, 5]. If this is the case, the system cannot 'know' the precise details of how heat was removed at such a remote distance. This means that the results obtained for the system using our simple mathematical thermostat must be the same as the those we would infer for the same system surrounded (at a distance) by a real physical thermostat (say with a huge heat capacity). This mathematical thermostat may be unrealistic, however in the final analysis it is very convenient but ultimately irrelevant devices. As Tolman pointed out [6], in a purely Hamiltonian system, the neglect of 'irrelevant' degrees of freedom (as in thermostats or by neglecting solvent degrees of freedom in a colloidal or Brownian system) will inevitably result in a non-conservation of phase space volume for the remaining 'relevant' degrees of freedom.

The exact equation of motion for the N-particle distribution function is the time reversible Liouville equation [1],

$$\frac{\partial f(\Gamma,t)}{\partial t} = -\frac{\partial}{\partial \Gamma}\bullet[\dot{\Gamma}f(\Gamma,t)] \equiv -iL(\Gamma)f(\Gamma,t) \qquad (3)$$

where $iL(\Gamma)$ is the distribution function (or $f$-) Liouvillean and appears in the propagator for the phase space distribution function ($f(\Gamma,t) = \exp(-iL(\Gamma)t)f(\Gamma,0)$). The Liouville equation can also be written in Lagrangian form [7],

$$\frac{df(\Gamma,t)}{dt} = -f(\Gamma,t)\frac{d}{d\Gamma}\bullet\dot{\Gamma} \equiv -\Lambda(\Gamma)f(\Gamma,t). \qquad (4)$$

This equation simply states that the time reversible equations of motion conserve the number of ensemble members, $N_\Gamma$. The presence of the thermostat is reflected in the phase space expansion factor, $\Lambda(\Gamma) \equiv \partial\dot{\Gamma}/\partial\Gamma$, which is to first order in N, $\Lambda = -d_C N_{therm}\alpha$. The equation of motion for an arbitrary phase function $B(\Gamma)$, is [1]

$$\dot{B}(\Gamma) = \dot{\Gamma}\bullet\frac{dB}{d\Gamma} \equiv iL(\Gamma)B(\Gamma). \qquad (5)$$

where $iL(\Gamma)$ is the phase variable (or $p$-) Liouvillean and appears in the propagator for phase variables ($B(\Gamma(t)) = \exp(iL(\Gamma)t)B(\Gamma(0))$). The difference between the $f$-Liouvillean and the $p$-Liouvillean is, $iL(\Gamma) - iL(\Gamma) = \Lambda(\Gamma)$.

The initial distribution can be written in a quite arbitrary form,

$$f(\Gamma,0) = \frac{\exp[-F(\Gamma)]}{\int d\Gamma \, \exp[-F(\Gamma)]}, \qquad (6)$$

where F is some arbitrary phase function. The Evans-Searles Transient Fluctuation Theorem [7-10] states that provided the system satisfies the condition of ergodic consistency [9], the dissipation function $\Omega(\Gamma)$, defined as [9, 10]

$$\int_0^t ds \, \Omega(\Gamma(s)) \equiv \ln\left(\frac{f(\Gamma(0),0)}{f(\Gamma(t),0)}\right) - \int_0^t \Lambda(\Gamma(s))ds$$
$$= \overline{\Omega}_t t \qquad (7)$$

satisfies the following time reversal symmetry [7-9]:

$$\frac{p(\overline{\Omega}_t = A)}{p(\overline{\Omega}_t = -A)} = \exp[At]. \qquad (8)$$

The TFT has generated much interest, as it shows how irreversibility emerges from the deterministic, reversible equations of motion, and is valid arbitrarily far from equilibrium. It provides a generalized form of the 2[nd] Law of Thermodynamics that can be applied to small systems observed for short periods of time. It also solves the longstanding Loschmidt Paradox. The TFT has been verified experimentally [12].

The *form* of the above equation applies to any valid ensemble/dynamics combination. However the precise *expression* for $\bar{\Omega}_t$ given in (7) is dependent on both the initial distribution *and* the dynamics. This result is extremely general. It is valid arbitrarily far from equilibrium. It leads to a number of other simple but important corollaries such as the Second Law Inequality [11],

$$\langle \bar{\Omega}_t \rangle \geq 0, \quad \forall t \tag{9}$$

and the NonEquilibrium Partition Identity,

$$\exp[-\bar{\Omega}_t t] = 1, \quad \forall t. \tag{10}$$

We now derive the *Dissipation Theorem*, which shows that the dissipation function is the central argument of both nonlinear and linear response theory (i.e. Green-Kubo theory). Following Evans and Morriss we can give a formal solution of the Liouville equation (3) using a Dyson decomposition of the *f*-propagator $\exp[-iL(\Gamma)t]$ in terms of the *p*-propagator $\exp[iL(\Gamma)t]$ (see §3.6 and equations (7.19, 20) of [1] for details), we find that,

$$\begin{aligned} f(\Gamma,t) &= \exp[-\int_0^t ds\, \Lambda(-s)] \exp[-iL(\Gamma)t] f(\Gamma,0) \\ &= \exp[-\int_0^t ds\, \Lambda(-s)] \exp[-F(\Gamma(-t)) - F(\Gamma(0))] f(\Gamma,0) \end{aligned} \tag{11}$$

Comparing this with equation (7) we see that

$$f(\mathbf{\Gamma},t) = \exp[-\int_0^{-t} ds\ \Omega(\mathbf{\Gamma}(s))]f(\mathbf{\Gamma},0). \tag{12}$$

Thus the propagator for the N-particle distribution function $\exp[-iL(\mathbf{\Gamma})t]$, has a very simple relation to exponential time integral of the dissipation function.

From equation (12) we can calculate nonequilibrium ensemble averages in the Schrödinger representation

$$\langle B(t)\rangle = \left\langle B(0)\exp[-\int_0^{-t} ds\ \Omega(\mathbf{\Gamma}(s))]\right\rangle, \tag{13}$$

and by differentiating and integrating (12) with respect to time, we can write the averages in the Heisenberg representation as

$$\langle B(t)\rangle = \int_0^t ds \langle \Omega(0)B(s)]\rangle. \tag{14}$$

In both sides of equations (13,14) the time evolution is governed by the field dependent thermostatted equations of motion (1). The derivation of (13,14) from the definition of the dissipation function (7), is called the *Dissipation Theorem*. This Theorem is extremely general. Like the Fluctuation Theorem it is valid arbitrarily far from equilibrium. As in the derivation of the Fluctuation Theorem the only unphysical terms in the derivation are the thermostatting terms within the wall region. However, because these thermostatting particles can be moved arbitrarily far from the system of interest, the precise mathematical details of the thermostat are unimportant. Since the number of degrees of freedom in the reservoir is assumed to be much larger than that of the system of interest, the reservoir can always be assumed to be in thermodynamic equilibrium. There is therefore no difficulty in defining the thermodynamic temperature of the walls. This is in marked contrast with the system of interest, which may be very far from equilibrium where the thermodynamic temperature cannot be defined.

For thermostatted dynamics where the kinetic energy $K_{therm}(\Gamma)$ of the thermostated particles is fixed and if the initial distribution is isokinetic

$$f(\Gamma,0) = \frac{\delta(K_{therm} - 3N_{therm}\beta^{-1})\exp[-\beta H_0(\Gamma)]}{\int d\Gamma\, \delta(K_{therm} - 3N_{therm}\beta^{-1})\exp[-\beta H_0(\Gamma)]}, \qquad (15)$$

$H_0(\Gamma)$ is the internal energy of the entire system, and $\beta = 1/(k_B T_W)$, it is straightforward [5] to show that the dissipation function is

$$\Omega(\Gamma) = -\beta \mathbf{J}(\Gamma) V \cdot \mathbf{F}_e . \qquad (16)$$

Here V is the volume of the system of interest and $\mathbf{J}(\Gamma)$ is the dissipative flux in the system of interest,

$$\mathbf{J}(\Gamma) V \cdot \mathbf{F}_e = -\sum_{i=1}^{N-N_W} [\frac{\mathbf{p}_i}{m} \cdot \mathbf{D}_i - \mathbf{F}_i \cdot \mathbf{C}_i] \cdot \mathbf{F}_e . \qquad (17)$$

Equation (14) can be written as the Transient Time Correlation function expression [1], for the thermostatted nonlinear response of the phase variable B to the dissipative field $\mathbf{F}_e$.

$$\langle B(t) \rangle = -\beta V \int_0^t ds \langle \mathbf{J}(0) B(s) \rangle \cdot \mathbf{F}_e . \qquad (18)$$

In the weak field limit this reduces to the well known Green-Kubo expression [1] for the linear response

$$\lim_{\mathbf{F}_e \to 0} \langle B(t) \rangle = -\beta V \int_0^t ds \langle \mathbf{J}(0)B(s)] \rangle_{\mathbf{F}_e=0} \cdot \mathbf{F}_e. \tag{19}$$

It is interesting to compare a number of different relationships between the distribution function, the dissipation function and the phase space expansion factor. The first such relation is equation (12) above. We note that although the time argument in (12) is negative, the dynamics must still be governed by the field dependent, thermostatted equations of motion (1). Rewriting (7) we have

$$f(\mathbf{\Gamma}(t),0) = \exp[-\int_0^t ds\, \Omega(\mathbf{\Gamma}(s)) + \Lambda(\mathbf{\Gamma}(s))]f(\mathbf{\Gamma}(0),0). \tag{20}$$

In a nonequilibrium steady state (SS), $\langle \Omega(t) \rangle_{ss} = -\langle \Lambda(t) \rangle_{ss}$. We also note that if the initial ensemble is microcanonical (has a uniform density) and the dynamics is isoenergetic, $\Omega(t) = -\Lambda(t), \quad \forall t$. Lastly we have the formal solution of the Liouville equation in its Lagrangian form (4),

$$f(\mathbf{\Gamma}(t),t) = \exp[-\int_0^t ds\, \Lambda(\mathbf{\Gamma}(s))]f(\mathbf{\Gamma}(0),0). \tag{21}$$

Rather obviously the results of the Dissipation Theorem (14) can also be used to obtain the Fluctuation Dissipation Theorem. Consider the case where the phase function $B(\Gamma) = J(\Gamma)$, and that the equilibrium dissipative flux autocorrelation function is δ-correlated, $\langle J(t_1)J(t_2)\rangle_{F_e=0} = \langle J(t_1)J(t_2)\delta(t_2-t_1)\rangle_{F_e=0}$, then we obtain the fluctuation dissipation relation,

$$\lim_{F_e \to 0} \langle J(t)\rangle = -\frac{1}{3}\beta V \langle J(0) \cdot J(0)]\rangle_{F_e=0} \cdot F_e .$$

In this paper we have shown the central importance of the dissipation function to nonequilibrium statistical mechanics. It is the argument of both the Fluctuation Theorem and the Dissipation Theorem. These Theorems are both exact arbitrarily far from equilibrium. The Fluctuation Theorem has been confirmed in laboratory experiments [12] and we see no reason why the Dissipation Theorem cannot be likewise tested in the laboratory.

Originally the dissipation function was defined in order to characterize the ratio of probabilities $p_r$, of observing infinitesimal bundles of phase space trajectories originating (t=0) in a volume $dV_{\Gamma(0)}$ to the probability of observing at t=0, their time reversed antitrajectories $dV_{\Gamma^*(0)}$, (Note: $\Gamma^*(0) \equiv M^T\Gamma(t)$),

$$\frac{p_r(dV_{\Gamma(0)},0)}{p_r(dV_{\Gamma^*(0)},0)} = \exp[\int_0^t ds\, \Omega(\Gamma(s))] . \qquad (22)$$

Combining (22) with (12) shows that the nonequilibrium N-particle distribution function at time t, can be written in terms of the ratio of probabilities of observing

$$f(\Gamma(0),t) = \frac{p_r(dV_{M^T\Gamma(-t)},0)}{p_r(dV_{\Gamma(0)},0)} f(\Gamma(0),0) . \qquad (23)$$

We find it remarkable that measure of *irreversibility* given in (22) by the dissipation function also features so centrally in the Dissipation Theorem. Our work shows that this measure of irreversibility is the prime function in determining how a nonequilibrium system will *respond* to a nonequilibrium perturbation or dissipative field.

We have given a derivation of the Dissipation Theorem for an exceedingly general set of time reversible equations of motion (1) and for an arbitrary initial distribution (6). We have argued that although the derivation employs unphysical thermostatting terms, these are just a convenient but ultimately irrelevant device. For both Theorems dissipation takes place only in the system of interest (as measured by the dissipation function) and on average, heat is lost to a surrounding reservoir region that is arbitrarily close to thermodynamic equilibrium at a known temperature.

It has been proposed that in nonequilibrium systems the path integral of the entropy production should be an extremum (see, for example, [13]). To leading order, close to equilibrium the dissipation function is the entropy production $\Sigma(t)$, $\lim_{\mathbf{F}_e \to 0}(\Omega(t) - \Sigma(t)) = O(\mathbf{F}_e)^4$. Our work shows that in natural systems that exchange heat with their surroundings, the maximum entropy production hypothesis, at best, can only be an approximation. In none of our expressions for the nonequilibrium N-particle distribution function (12, 20, 21) is the probability density controlled solely by the path integral of the entropy production.

**ACKNOWLEDGEMENTS**

We would like to thank Lamberto Rondoni for his valuable comments on this work. We also thank the Australian Research Council for its support of this work.